\documentclass[sigconf,screen]{acmart} 

\AtBeginDocument{%
  \providecommand\BibTeX{{%
    \normalfont B\kern-0.5em{\scshape i\kern-0.25em b}\kern-0.8em\TeX}}}

\setcopyright{rightsretained}
\acmDOI{}

\acmConference[CHI '24 Sensemaking Workshop]{Sensemaking Workshop, CHI Conference on Human Factors in Computing Systems}{May 11--16, 2024}{Honolulu, HI, USA}
\acmBooktitle{Sensemaking Workshop, CHI Conference on Human Factors in Computing Systems (CHI '24), May 11--16, 2024, Honolulu, HI, USA}
\acmISBN{}

\usepackage{todonotes}
\usepackage{enumitem}

\begin{document}

\title{Tools and Tasks in Sensemaking: \\ A Visual Accessibility Perspective }


\author{Yichun Zhao}
\affiliation{%
  \institution{University of Victoria}
  \city{Victoria BC}
  \country{Canada}}
\email{yichunzhao@uvic.ca}

\author{Miguel A. Nacenta}
\affiliation{%
  \institution{University of Victoria}
  \city{Victoria BC}
  \country{Canada}}
\email{nacenta@uvic.ca}



\begin{abstract}
Our previous interview study explores the needs and uses of diagrammatic information by the Blind and Low Vision (BLV) community, resulting in a framework called the \textit{Ladder of Diagram Access}. The framework outlines five levels of information access when interacting with a diagram. In this paper, we connect this framework to include the global activity of sensemaking and discuss its (in)accessibility to the BLV demographic. We also discuss the integration of this framework into the sensemaking process and explore the current sensemaking practices and strategies employed by the BLV community, the challenges they face at different levels of the ladder, and potential solutions to enhance inclusivity towards a data-driven workforce. 
\end{abstract}




\maketitle


\section{Introduction}

Sensemaking is the process of creating and refining understanding from complex and uncertain data and it is a crucial skill for many professional domains~\cite{Jolaoso_Burtner_Endert_2015}. However, sensemaking often relies heavily on visual representations such as diagrams, charts, and graphs, to convey and manipulate information~\cite{Russell_2003}. This poses a significant challenge for people with visual disabilities, who may have limited or no access to these representations.  

In this paper, we explore the topic of sensemaking from a visual accessibility perspective and discuss the current practice in this area. We believe diagrammatic information holds a unique and compelling interest rooted in the seminal work of Larkin and Simon, who argued that diagrams can aid problem-solving and cognitive tasks by reducing search, supporting recognition over recall, and facilitating perceptual inference~\cite{Larkin_Simon_1987}. Diagrams, by their very nature, present information in a spatially organized manner that allows for efficient processing and interpretation to represent complex relationships and structures more intuitively than textual or numerical data, making them a powerful tool for sensemaking. We delve into the insights derived from a qualitative interview study conducted by Zhao et al.~\cite{Zhao_2023, Zhao_Nacenta_Sukhai_Somanath_2024}. The study involved 15 blind and low-vision (BLV) individuals who shared their experiences and challenges with accessing and using diagrams in various contexts. These insights are encapsulated in the \textit{Ladder of Diagram Access}, a framework that characterizes the different levels of information access that BLV people can achieve when interacting with a diagram, and the factors that influence their access. 


We then apply this framework to the sensemaking process and show how it can help us understand and address the accessibility barriers that BLV people face when performing complex, professional tasks that involve data analysis and visualization. We also discuss the potential impact of our findings on the design of tools and tasks for sensemaking and the implications for creating a more inclusive and diverse data-driven workforce. Finally, we acknowledge the limitations of our study and suggest directions for future research. Our contribution is not a reiteration of these findings, but rather a reinterpretation and application of them through the lens of sensemaking. We explore how the \textit{Ladder of Diagram Access} intersects with the sensemaking process, providing a fresh perspective on the original study.

\section{Methods}

We revisit a qualitative interview study conducted with 15 BLV participants, all of whom had varying degrees of vision loss and experience with diagrams. This study was originally undertaken as part of a larger project aimed at designing and evaluating a novel tool for making node-link diagrams accessible to BLV individuals~\cite{Zhao_2023, Zhao_Nacenta_Sukhai_Somanath_2024}. The study followed a semi-structured interview protocol, asking participants to describe their encounters with diagrams in various contexts and demonstrate their preferred methods of accessing and using diagrams. The open coding and thematic analysis of the interview transcripts led to the emergence of 124 codes, which were grouped into five themes. These themes collectively formed the \textit{Ladder of Diagram Access} framework, a significant outcome of the study. In this paper, we further delve into the codebook, reanalyzing these insights to better align with the data sensemaking process by adopting established models and theories of sensemaking. 


\section{Results: Ladder of diagram Access}



This framework describes the five levels of information access that BLV people can achieve when interacting with a diagram, and the factors that influence their access. From the lowest to the highest: 

\begin{enumerate}[leftmargin=16pt]
\item \textbf{Unaware of the Representation}: This bottom level corresponds to the situation where BLV individuals are unaware of a diagram’s existence due to it being skipped or ignored by their assistive technology or informants, having no access to the data. 

\item \textbf{Aware of the Representation}: BLV individuals know a diagram exists but lack further information about it. For example, they may hear a placeholder such as “graphic” or “embedded object” from their screen reader, but they do not know what type of diagram it is, what it represents, or why it is important. They know the existence of the data, but not to the data itself or the schema of the diagram. 

\item \textbf{Single Static Perspective}: BLV individuals learn about the diagram from a single static perspective (e.g., alternative text), but lack control over the level of detail and granularity of the information.

\item \textbf{Multiple Perspectives}: BLV individuals gain additional perspectives of the diagram beyond a static description, allowing more interactive and flexible query and navigation. This supports different tasks and goals that BLV people may have when accessing a diagram, such as finding relations, summarizing, or reflecting.  

\item \textbf{Comprehensive Access}: The top level corresponds to the ideal situation where BLV individuals access the diagram comparably to sighted individuals in terms of efficiency and effectiveness, enabling tasks like understanding, analyzing, and creating new knowledge from the data.
\end{enumerate}

\subsection{Current Practices and Challenges related to Sensemaking}



Within the scope of the previous interview study, we have learnt about participants' current experiences with accessing and understanding diagrammatic information at different levels on the ladder. At the lower levels 1 and 2, they often struggle to know the existence of a diagram, to understand its purpose and content, and to get a single static perspective of it. Four out of 15 participants resort to using OCR or CV tools to scan and read diagrams to move up to level 3 but accuracy and reliability cannot be ensured. Six seek information through alternative means but this takes time and efforts. To move up to level 4 and 5, eleven participants seek additional and complimentary information from other sources providing difference perspectives. Many (14 out of 15) rely on sighted individuals to describe diagrams and can engage in a back-and-forth conversation to query the diagram. 


However, several barriers to sensemaking emerge. The success of sensemaking depends on individual needs (n=5), and a static representation is limited as P12 explains: ``Sometimes, you just want to skim something, and you can’t. You can’t do the “cheaty” short[cut] method of looking at the diagrams [visually] and inferring the information.” P6 highlights that the single perspective might not be enough for tasks which require more information: ``I don’t have enough information [from the diagram] ... I think it makes a difference too if I actually need it for something [to perform tasks on], or if it’s just kind of something that’s just there.'' P10 also confirms that it is ``a big challenge'' to match the different goals of individuals, such as the diagram ``design'', ``layout'', or ``statistical information''. 

With multiple perspectives, there could be a lack of consistency in the information accessed from them (n=4). P11 explained: “The challenge is ... [to] work out [multiple accesses to information] that are going to work universally [so that] I, the next person, and the next person after will be able to interpret them all in a similar way.” P10 added that if there is inconsistency, then they would ``[start to] invest time in trying to understand that. ''
P8 had to sometimes combine multiple perspectives from other people: “Everybody has a different way of explaining something, and if somebody missed some key things, then it’s important to get an explanation from 2 or 3 people, and then I can put things together better.” P14 highlighted that crucial information might be missed when transitioning between an overview of a diagram and its specific details and gave an example of occasionally missing connections between individual diagram entities. Five participants also reported the need to iteratively assess the relevance of information currently presented and switch perspective if necessary. 

Despite these challenges to sensemaking, participants shared strategies to overcome them. One approach is to divide the information into digestible parts or layers (n=3): ``It’s always best to simplify and break up the topic into multiple diagrams, or you risk making it useless” [P2]. For tactile diagrams, P11 use different textures to distinguish various elements and regions. Additionally, 14 people recommended to consider both the overview and the details of a diagram, as confirmed by P8: ``Start with a basic outline and then try filling in the detail.'' Two participants would rearrange spatial layouts mentally to enhance sensemaking: ``Sometimes you have to come up with new ways of representing things ... so that you know this relates to this, but in an alternative way it relates to that as well” [P7]. 13 participants also annotate alongside the diagrams. Participants at level 4 or 5 can conduct more sensemaking-related tasks such as cross-referencing and comparing (n=10), finding patterns or commonalities (n=5), summarization (n=1) and reflection (n=1).

\section{Discussion: Sensemaking and Visual Accessibility }

We elaborate on the implications that the challenges BLV people face when accessing diagrams have on sensemaking. Multiple perspectives sometimes lead to inconsistencies and confusion, cause people to struggle with reconciling these conflicting details affecting the audience's understanding of the data. The Iceberg Sensemaking model marks the importance of considering both explicit and tacit schemas when interpreting data~\cite{Berret_Munzner_2023}, and any inconsistencies between these schemas can obstruct the sensemaking process. As a result, people may find themselves spending more time synthesizing information from various sources, which is also part of the iterative development and refinement of knowledge structures~\cite{Jolaoso_Burtner_Endert_2015, Pirolli_Card_2005}. However, we learnt from the interview study that this cost is even more significant for BLV people. Therefore, it is important to strike a balance between perspectives to manage cognitive load. Lastly, the reliance on sighted individuals or other sources for additional perspectives can also be limiting, with availability, expertise, and social dynamics influencing the quality and timeliness of these perspectives. The Iceberg Sensemaking model acknowledges this complexity and highlights the role of power dynamics and the need to articulate tacit knowledge explicitly~\cite{Berret_Munzner_2023}.



The ladder of access can help us understand how the sensemaking loop and the data frame~\cite{Jolaoso_Burtner_Endert_2015} model can be used more effectively to transform data into knowledge and insights. The sensemaking process is curtailed at the lower levels in the ladder, where individuals may struggle to know the existence of a diagram, understand its purpose and basic content and assess its relevance. By moving up the ladder, people can leverage multiple perspectives to support and enhance their sensemaking process. In fact, sensemaking tools commonly assume that users are already at level 4 and able to access additional perspectives. The higher levels in the ladder also emphasize the iterative nature of sensemaking, where people develop and refine their mental models through activities like foraging and synthesis~\cite{Russell_Stefik_Pirolli_Card_1993, Jolaoso_Burtner_Endert_2015}. Multiple and interactive perspectives of information also allow BLV people to form alternative representations which are part of the learning loop complex~\cite{Russell_Stefik_Pirolli_Card_1993}. For example, a BLV project manager who wants to make sense of a Gantt chart diagram may be at level 2 if they only know that there is a graphic on the web page, but not what it represents. They may move to level 3 if they can access a static description of the diagram, such as an alt-text or an OCR output. However, this may not be sufficient for their sensemaking needs, as they may want to explore the diagram in more detail, compare different diagram elements, or test different hypotheses. To achieve level 4, they may need to access multiple perspectives of the diagram, such as a summary, an overview, or a query. This may require the help of sighted people or the use of specialized tools like tactile graphics. To reach level 5, they may need to access the diagram in an efficient and effective way, without excessive cognitive or social costs, and ideally with the same benefits as sighted people. 

Representations like diagrams are spatial in nature and can take advantage of visual processing~\cite{Russell_2003}. We have learnt from our BLV participants that spatial information is crucial in making sense of information to understand the “items” in a diagram, the “relationships of space and the items that take up the space”, and the “space in between them” [P2]. Although the visual aspect of spatial information might not be perceived successfully, one can still understand spatial information by relying on other sensory inputs. One example given by our participants was that they could identify salient features of a tactile diagram by distinguishing different tactile textures or symbols. Considering alternative representations beyond visual ones is therefore important. At the same time, we also learnt that an representation should afford interactions that allows the audience to filter information in different layers to help them construct the mental models of the information more effectively. 

The ladder of diagram access can be integrated into the sensemaking process, as it affects how BLV people gather, represent, and communicate information. The higher the level of access, the more likely BLV people can perform sensemaking tasks, such as finding patterns, generating insights, and creating solutions. However, the level of access may also depend on the sensemaking context, such as the task, the domain, and the goal~\cite{Jolaoso_Burtner_Endert_2015}. For example, a BLV person who wants to make sense of a networking diagram may only need level 3 access if they are interested in the overall distribution of the data, but they may need level 4 or 5 access if they want to compare or drill down into the details.

\subsection{Towards an Inclusive Data-driven Workforce}

Despite the difficulties, our interview study confirmed that many BLV people are eager and capable of participating in the data-driven workforce and that they have developed a variety of coping strategies and workarounds to overcome accessibility barriers to sensemaking. These findings also have implications for the design of tools and interventions that can support sensemaking for BLV people, and contribute to an inclusive data-driven workforce. Some of the potential solutions and strategies for addressing the identified challenges include: 

\paragraph{Universal Design} This refers to the design of products, environments, and systems that are accessible and usable by all people, regardless of their abilities, preferences, or contexts~\cite{Lid_2014}. Universal design can help BLV people achieve level 5 access by providing multiple modalities to represent and interact with diagrams. For example, a universal design tool could allow BLV users to switch between different modes of presentation, such as text, speech, sound or vibration, depending on their needs and preferences. Universal design can also benefit sighted people, as it can enhance their cognitive and perceptual abilities, and accommodate their diverse learning styles and situations~\cite{Mayer_2002}. 

\paragraph{Translation Tools} These are tools that can translate information from one representation to another, such as from diagrammatic to linear beyond the modality. Translation tools can help BLV people achieve level 4 or 5 access by providing multiple perspectives of information and allowing them to explore and manipulate diagrammatic information in their preferred modality. For example, a translation tool could convert a bar chart into a series of tones, where the pitch, volume, and duration of each tone correspond to the height, width, and label of each bar. By providing alternative ways of perceiving and understanding diagrams, translation tools can also potentially reveal hidden or complex patterns and relationships. 

\paragraph{Sensemaking Tools} These support the sensemaking process, such as finding, framing, filtering, organizing, analyzing, and communicating information~\cite{Russell_Stefik_Pirolli_Card_1993, Russell_2003, Jolaoso_Burtner_Endert_2015, Weick_Sutcliffe_Obstfeld_2005}. Sensemaking tools can help BLV people achieve level 5 access by providing features and functions that can reduce the cognitive and social costs of accessing diagrams, and enhance the benefits of generating insights and solutions. For example, a sensemaking tool could provide a summary, an overview, or a query of the diagram, as well as note-taking, hypothesis-testing, or a collaboration function. Sensemaking tools can involve design principles from universal design and translation tools to ultimately facilitate information processing and decision-making.







\section{Limitations and Future Directions}

While our study provides valuable insights into the challenges and strategies of BLV people when accessing and using diagrams, it also has several limitations. Our study focused on one type of representation of information being diagrammatic, so our findings may not apply to other types of representations. In future research, we plan to conduct more in-depth studies to investigate the specific challenges and strategies of BLV people when performing sensemaking tasks in a professional setting and also in a collaborative setting. We believe our work can contribute to the ongoing efforts to empower all people to participate fully in the data-driven workforce. 


\bibliographystyle{ACM-Reference-Format}
\bibliography{sample-base}


\end{document}